\documentclass[journal]{IEEEtran}

\usepackage{cite}
\usepackage{amsmath,amssymb,amsfonts}
\usepackage{algorithmic}
\usepackage{graphicx}
\usepackage{textcomp}
\usepackage{xcolor}
\usepackage{bm}
\usepackage{amsthm}
\usepackage[utf8]{inputenc}
\usepackage[T1]{fontenc}
\usepackage{lmodern}
\usepackage[english]{babel}

\newtheorem{definition}{Definition}

\def\BibTeX{{\rm B\kern-.05em{\sc i\kern-.025em b}\kern-.08em
    T\kern-.1667em\lower.7ex\hbox{E}\kern-.125emX}}

\begin{document}

\title{QADR: A Scalable, Quantum-Resistant Protocol for Anonymous Data Reporting}

\author{Nilesh Vyas and Konstantin Baier%
\thanks{N. Vyas is with Airbus Central R\&T, Taufkirchen, 82024 Germany (e-mail: nilesh.vyas@airbus.com).}%
\thanks{K. Baier is with the Hochschule München für angewandte Wissenschaften, Munich, 80335 Germany. (e-mail: baierkonstantin0@gmail.com)}%
}


\maketitle

\begin{abstract}

The security of future large-scale IoT networks is critically threatened by the ``Harvest Now, Decrypt Later'' (HNDL) attack paradigm. Securing the massive, long-lived data streams from these systems requires protocols that are both quantum-resistant and highly scalable. Existing solutions are insufficient: post-quantum classical protocols rely on computational assumptions that may not hold for decades, while purely quantum protocols are too resource-intensive for the sheer scale of IoT.  This paper introduces the Quantum Anonymous Data Reporting (QADR) protocol, a hybrid framework that provides a theoretical benchmark and high-performance architecture for this challenge, designed for future fully-connected quantum networks. The protocol achieves scalable, quantum-resistant anonymity through a hybrid security model; it leverages information-theoretically secure keys from Quantum Key Distribution (QKD) to seed a quantum-secure pseudorandom function (QS-PRF), grounding its long-term data protection in well-established computational hardness assumptions. We also propose and analyze an automated slot reservation mechanism by making a deliberate trade-off: achieving high performance by accepting a quantifiable information leak during the anonymous slot reservation phase while maintaining strong unlinkability for the final data submission.  Our security analysis formally quantifies the anonymity reduction caused by the leak and discusses pathways to fully mitigate it at a significant performance cost.
We prove the protocol's critical advantage as a performance benchmark: its primary communication cost scales as $O(n^2)$, a dramatic improvement over quantum-native alternatives ($O(n^4)$), establishing a high-performance goal for future quantum-secured anonymity systems.
\end{abstract}


\section{Introduction}
\IEEEPARstart{T}{he} proliferation of mobile and IoT devices has catalyzed the emergence of large-scale participatory sensing systems—\textit{i.e., applications where data is collected from a large group of individuals via mobile devices, such as traffic data from smartphones or public health reports}— which now underpin critical applications ranging from public health surveillance to intelligent transportation networks \cite{Burke2006ParticipatorySensing}. While these systems generate unprecedented value, they do so by collecting data streams that are intrinsically tied to individuals, creating profound privacy risks \cite{Christin2011Privacy}. Even when encrypted, associated metadata can be subjected to powerful traffic analysis techniques, compromising participant anonymity \cite{Danezis2015Traffic}.

This privacy challenge is magnified by the impending arrival of fault-tolerant quantum computing. The public-key cryptosystems (e.g., RSA, ECC) that form the bedrock of modern digital security are rendered insecure by quantum algorithms like Shor's \cite{Shor1997Polynomial}. This vulnerability enables the ``Harvest Now, Decrypt Later'' (HNDL) attack paradigm: adversaries can intercept and archive today's encrypted data with the express intent of decrypting it once a cryptographically relevant quantum computer is operational \cite{Bernstein2017PostQuantum}. For participatory sensing data, which often has long-term value, HNDL demands security solutions that offer multi-decade, post-quantum guarantees.

Existing frameworks are ill-equipped to meet this dual need for scalability and enduring security. Classical anonymity networks like Tor are vulnerable to quantum adversaries \cite{tor}. Hardening classical protocols like DC-Nets with Post-Quantum Cryptography (PQC) provides an important near-term defense, but this approach still tethers long-term security to computational assumptions that could theoretically be broken by future algorithmic breakthroughs \cite{kyber}. On the other end of the spectrum, purely quantum anonymity protocols like Anonymous Private Message Transmission (APMT) \cite{broadbent2007, Huang2022} offer strong security guarantees but are often highly interactive and complex, rendering them fundamentally unscalable for the bulk reporting required by IoT ecosystems with thousands or millions of endpoints. This leaves a critical gap: a need for a protocol that is both practical for massive networks and secure for the long term.

This paper proposes the Quantum Anonymous Data Reporting (QADR) protocol, a flexible hybrid framework designed for scalable and secure data reporting in the post-quantum era. The novelty of QADR is not a mere substitution of cryptographic primitives but a ground-up redesign for genuine quantum resistance. We demonstrate that classical Bulk Transfer Protocol (BTP) architectures~\cite{li2021}, while scalable, rely on mechanisms like homomorphic encryption that are broken by quantum computers, necessitating a new approach.

Our work delivers a comprehensive solution by presenting QADR as a framework with two distinct mechanisms for anonymous slot reservation, offering a clear trade-off between performance and security.
First, we introduce a novel, high-performance protocol for dynamic slot reservation and provide the first formal analysis of a critical, yet previously unaddressed, information leak inherent to this scalable architecture. We prove that despite this leak partitioning the anonymity set, the core property of unlinkability is preserved.
Second, for scenarios demanding maximum security, we present a robust alternative based on a verifiable oblivious shuffle. This mechanism eliminates the information leak entirely and provides stronger guarantees against active adversaries, establishing a new benchmark for quantum-resistant anonymity.

The result is a complete, synergistic protocol that leverages information-theoretically secure keys from Quantum Key Distribution (QKD)~\cite{Bennett2014QuantumCrypto} as its root of trust while harnessing the BTP's efficiency for data aggregation. Our contributions are threefold:

\begin{enumerate}
    \item We present the complete QADR framework, architected for parallel data submission, detailing its two distinct mechanisms for anonymous slot reservation: a high-throughput, collision-managed protocol and a high-security variant using a verifiable oblivious shuffle.
    \item We provide a formal security analysis for both mechanisms. We quantify the information leak in the high-throughput mode while proving its preservation of unlinkability, and we demonstrate how the shuffle-based mode provides stronger anonymity against active quantum adversaries.
    \item We conduct a rigorous performance evaluation demonstrating that QADR's core data submission complexity scales as $O(n^2)$, a fundamental gain over comparable quantum protocols scaling at $O(n^4)$, and we analyze the distinct cost profiles of our two proposed reservation schemes.
\end{enumerate}
It is important to state at the outset that QADR is designed as a forward-looking protocol for a next generation of quantum infrastructure. Its security guarantees assume the existence of a fully-connected, peer-to-peer quantum network where any two participants can establish direct QKD links. While this technology is not yet mature, our work aims to establish a robust and highly scalable algorithmic framework so that privacy solutions are ready for deployment as the underlying hardware evolves. 




\section{\label{sec:level2}Related work}
The pursuit of anonymity has produced a spectrum of solutions with varying trade-offs between security, efficiency, and scalability.

\subsection{Classical Anonymous Communication}
Foundational work by Chaum introduced mixnets \cite{chaum1981}, which use a series of proxy servers to reorder and decrypt messages in batches, thereby obscuring sender-receiver links. While effective, the batching process introduces high latency, making them unsuitable for many applications. Chaum also proposed DC-nets (Dining Cryptographers Networks) \cite{chaum1988dc}, which offer information-theoretic anonymity against traffic analysis by broadcasting messages as the XOR sum of inputs masked with shared secret key \cite{wacksman2004}. However, traditional DC-nets suffer from high communication overhead and a vulnerability to jamming attacks \cite{wacksman2004}. The most widely deployed low-latency system is Tor, based on Onion Routing\cite{tor}. It provides scalability for millions of users but is vulnerable to traffic correlation attacks by a global adversary and relies on classical cryptography that is not secure against future quantum attacks.

\subsection{Post-Quantum Classical Anonymity}
A direct approach to quantum resistance is to replace the classical cryptographic components of existing anonymity systems with PQC algorithms. For instance, the pairwise keys in a DC-Net can be established using a PQC key encapsulation mechanism (KEM) like CRYSTALS-Kyber \cite{kyber}. While this provides resistance to known quantum algorithms, its security is still based on computational assumptions (e.g., the hardness of the Learning with Errors problem). A future breakthrough in classical or quantum algorithms could compromise these systems. QADR's hybrid approach, using QKD for the root keys, provides a stronger foundation, as the key secrecy is guaranteed by physical principles, not computational hardness.

\subsection{Quantum Anonymous Communication}
Quantum mechanics offers primitives for information-theoretically secure anonymity. Anonymous Quantum Conference Key Agreement (AQCKA) allows a group to establish a shared key while concealing identities \cite{AQCKA, christandl2005}, though some protocols require complex multipartite states, while others can be built from bipartite entanglement \cite{grasselli2022anonymous}. A more direct benchmark is the \textit{Anonymous Private Message Transmission} (APMT) protocol, demonstrated on an eight-user quantum network \cite{Huang2022}. Each user is connected with each other and shares pairwise secret keys, which can be established using Quantum Key Distribution (QKD). The protocol is based on the Dining Cryptographers Problem~\cite{wacksman2004} and consists of the following five steps: Anonymous Broadcasting, Veto, Notification, Collision Detection, and Anonymous Private Message Transmission. Each subprotocol's information-theoretic security was proven by Broadbent and Tapp~\cite{broadbent2007}. However, its highly interactive and serial nature, requiring multiple rounds of communication for a single message, results in a resource cost that scales poorly for multi-user bulk reporting, as we demonstrate in Section \ref{sec:performance}.

\section{Preliminaries}
\label{Prelim}

\subsection{Bulk Transfer Protocol (BTP)}
The BTP \cite{li2021} is a scalable protocol derived from DC-Net principles for anonymously collecting messages from $n$ participants. The protocol assumes that every pair of participants ($P_i, P_j$) shares a secret seed $S_{ij}$ and knows their pre-assigned, secret submission slot. Each participant $P_i$ constructs a vector $C_{i}^{\text{msg}}$ containing their message $\mathrm{msg}_i$ in their slot, padded with zeros. They then mask this vector by XORing it with pseudorandom pads $C_{ij}$ generated from the shared seeds $S_{ij}$.
\begin{equation}
    C_{i} = C_{i}^{\text{msg}} \oplus \bigoplus_{j \neq i} C_{ij}
\end{equation}
Each participant sends their final vector $C_i$ to a server. The server XORs all received vectors. Since each pad $C_{ij}$ is added twice, they cancel out, leaving the sum of the message vectors: $\bigoplus C_i = \bigoplus C_i^{\text{msg}}$. The server obtains a concatenation of all messages but cannot determine the origin of any individual message. 

While this data submission stage is agnostic to the key source, a critical prerequisite is the anonymous assignment of slots. The mechanism proposed in \cite{li2021} to solve this challenge is specifically built on classical primitives (Paillier homomorphic encryption), which are insecure against quantum adversaries. Therefore, to make the BTP architecture viable in a post-quantum setting, a completely new, quantum-resistant mechanism for anonymous slot assignment is required. Designing this mechanism is a core contribution of QADR.

\subsection{Quantum-Secure Cryptographic Primitives}
Our protocol's security relies on a hybrid cryptographic approach.

\textit{Quantum Key Distribution (QKD)} allows two parties to establish a shared secret key with information-theoretic security, guaranteed by the principles of quantum mechanics \cite{Bennett2014QuantumCrypto}. In our protocol, QKD is used to generate pairwise secret seeds ($S_{ij}$), which form the root of trust.

\textit{Quantum-Secure Pseudorandom Functions (QS-PRF)} are designed to be indistinguishable from a truly random function, even against an adversary with quantum query capabilities \cite{BPR12}. We employ a QS-PRF as a cryptographic extender. Standard constructions for QS-PRFs are based on problems believed to be hard for quantum computers, such as Learning With Errors (LWE).

The combination of these primitives represents a crucial design trade-off. While the key generation rate of QKD has historically been a limitation, point-to-point key rates for commercial systems now reach kilobytes per second at a metropolitan scale \cite{Madqci}. This rate is sufficient for many participatory sensing applications that involve sending small data payloads, such as traffic congestion alerts, public transit occupancy reports, or simple environmental sensor readings. In these scenarios, the protocol can use the QKD keys directly as one-time pads, achieving full information-theoretic security.

For applications requiring the transfer of large data, such as  submitting high-resolution images or raw scientific data streams, we employ a QS-PRF as a cryptographic extender for practical scalability. In these cases, the overall security of the QADR protocol is not information-theoretic but rests on the computational hardness assumptions of the chosen QS-PRF against quantum adversaries. This hybrid approach provides robust, long-term security suitable for protecting against HNDL attacks while remaining practical across a wide spectrum of real-world applications.

\section{System and Threat Model}
\label{sec:system_model}

\subsection{System Architecture}
The system consists of three main entities:
\begin{enumerate}
    \item \textbf{Task Requester (TR):} The trusted entity that commissions the data collection task and is the final recipient of the data.
    \item \textbf{Server/Service Provider (SP):} An intermediary that manages tasks and aggregates data. The SP is considered \textit{honest-but-curious}, meaning it follows the protocol correctly but may attempt to de-anonymize users.
    \item \textbf{Participants (P):} A set of $n$ users $\{P_1, \dots, P_n\}$ who collect and report data. We assume that within any group, at least two participants are honest and do not collude.
\end{enumerate}
Participants establish pairwise QKD keys ($S_{ij}$) and submit masked data to the SP. The SP aggregates the data and forwards it to the TR.

The protocol's full security guarantees are designed for an environment where any two participants can establish pairwise QKD keys directly requiring a  fully-connected peer-to-peer quantum networks, a goal toward which quantum communication technology is rapidly advancing. While current quantum communication often utilizes trusted-node architectures \cite{trusted_node, Madqci}, which have been successfully deployed in metropolitan areas, this paper's focus is on providing a forward-looking solution. It presents a scalable, quantum-resistant anonymity protocol intended for the next generation of quantum infrastructure. The primary contribution is the development of a robust algorithmic framework that can be integrated into future networks as they mature, ensuring that scalable privacy solutions are ready for widespread deployment.


\begin{figure}[!t]
    \centering
    \includegraphics[width=\columnwidth]{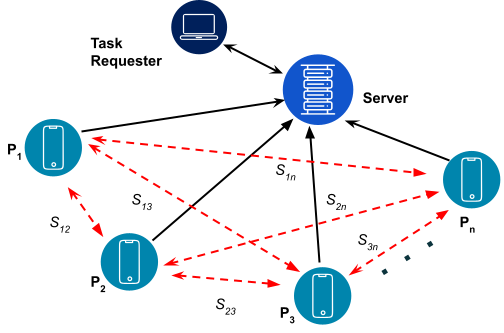}
    \caption{System Architecture. Participants ($P_i$) establish pairwise QKD keys ($S_{ij}$) and submit masked data to the Server (SP).}
    \label{fig:system}
\end{figure}

\subsection{Threat Model}
\label{ThreatModel}
We consider a global, active adversary operating under the Harvest Now, Decrypt Later (HNDL) attack paradigm. The adversary can monitor all communications, corrupt up to  $n-2$ participants, and may be the Service Provider (SP) itself. Their strategy is to intercept and archive all protocol communications today. Their ultimate goal is to use a future quantum computer to break the protocol's cryptographic protections, allowing them to retroactively deanonymize the honest participants by decrypting the stored data and linking them to their past submissions.

\subsection{Security Goals}
The QADR protocol is designed to achieve the following security goals against the adversary defined above:
\begin{definition}[Sender Anonymity]
The adversary cannot determine the identity of the originator of a specific message among the set of honest participants.
\end{definition}
\begin{definition}[Unlinkability]
The adversary cannot link a participant's identity to their submitted message, even when given a set of participant identities and a set of submitted messages. The adversary's view of the protocol transcript should be statistically indistinguishable regardless of the permutation mapping honest participants to their messages.
\end{definition}

\section{Quantum Anonymous Data Reporting (QADR) Protocol}
\label{QADR}
The Quantum Anonymous Data Reporting (QADR) protocol is our proposed framework for secure and anonymous data reporting in participatory sensing environments. It is built upon two core components: a classical \textit{Bulk Transfer Protocol (BTP)} for data aggregation and \textit{Quantum Key Distribution (QKD)} for establishing shared secrets. 

\subsection{Setup Stage}
The setup stage establishes the necessary cryptographic foundation and operational parameters for a group of participants. This initial process is crucial for the secure operation of all subsequent protocol phases.

\begin{enumerate}
    \item \textit{Registration:} New participants must first register with a the Service Provider (SP), providing their real-world identity. 
    \item \textit{Quantum Key Establishment:} Once a group of $n$ participants ($P_1, P_2, \dots, P_n$) is formed, they establish pairwise symmetric keys with every other member of the group. This is accomplished using Quantum Key Distribution (QKD), which generates a unique, information-theoretically secure secret key $S_{ij}$ for each pair of participants ($P_i, P_j$). 
    \item \textit{Parameter Agreement:} Finally, all participants in the group and the SP collectively agree on the operational parameters for the protocol. This includes defining the standard format and bit-length of the Slot Reservation Messages ($l_{\text{srm}}$), the QS-PRF to be used, and determining the total number of available slots ($m$) to be used during the slot reservation phase.
\end{enumerate}
Although we present the protocol with QS-PRFs for scalability to meet the requirements for sending large data, one can use the ITS secure QKD keys directly for applications that only require sending small data packets.

With the successful completion of this setup stage, the participants are equipped with the shared secrets and common parameters required to proceed to the \textit{slot reservation} and \textit{message transmission} stages of the protocol.

\subsection{Slot Reservation Stage}
The slot reservation stage anonymously assigns a unique data submission position to each participant. This is achieved by applying the principles of the bulk transfer protocol, but reinforced with quantum-secure cryptography. The final output is a vector where participants' requests appear in a jumbled order, concealing who requested which slot.

The process involves three main steps: participant-side preparation, aggregation by the Service Provider (SP), and collision resolution.

\subsubsection{Participant Actions}
Each of the $n$ participants ($P_i$) executes the following steps to prepare their slot reservation request:
\begin{itemize}
    \item  Each participant $P_i$ generates a fresh, random secret pseudonym, $\mathit{PN}_i$, of $\lambda$ bits. This pseudonym acts as a unique, verifiable identifier for this session only.
    \item The participant creates a \textit{Slot Reservation Message}, $\text{SRM}_i$. Its format and length, $l_{\text{srm}}$, are pre-agreed upon. The message contains at least the pseudonym, $\text{SRM}_i = (\mathit{PN}_i, \dots)$.
    \item   $P_i$ randomly selects one of the $m$ available slots and constructs a vector, $C_{i}^{\text{srm}}$, of total length $L_v = m \times l_{\text{srm}}$. This vector contains $\text{SRM}_i$ in the chosen slot and is padded with zeros elsewhere.
    \begin{equation}
        C_{i}^{\text{srm}} = \underbrace{\overbrace{0 \cdots 0}^{\text{preceding slots}} | \text{SRM}_i | \overbrace{0 \cdots 0}^{\text{succeeding slots}}}_{L_v \text{ total bits}}
    \end{equation}
    \item Next, using the secret keys $S_{ij}$ established via QKD, $P_i$ generates $n-1$ quantum-secure pseudo-random bitstreams, $C_{ij}$.
    \begin{equation}
        C_{ij} = \text{QS-PRF}\{L_v, S_{ij}\}, \quad \text{for } j \in  \{1,\cdots, n \}, j \neq i
    \end{equation}
    \item Finally, $P_i$ computes its final vector, $C_i$, by XORing its message vector with all the quantum-secure pads.
    \begin{equation}
        C_{i} = C_{i}^{\text{srm}} \oplus \bigoplus_{j \neq i} C_{ij}
    \end{equation}
\end{itemize}
\subsubsection{Service Provider (SP) Aggregation}
Each participant transmits their vector $C_i$ to the Service Provider. The SP performs a single operation: XORing all received vectors together. Because each masking pad $C_{ij}$ is added twice (once by $P_i$ and once by $P_j$), they cancel each other out.
\begin{equation}
    C  = C_{1} \oplus C_{2} \oplus \cdots \oplus C_{n} = \bigoplus_{i=1}^{n} C_{i}^{\text{srm}}
\end{equation}
The resulting public vector $C$ contains the SRMs from all participants, distributed among the $m$ available slots in a permuted order, with empty slots filled with null bits (`\#').
\begin{equation}
    C = \underbrace{\#|\cdots| \mathrm{SRM}_{\pi(1)}|\cdots| \#|\cdots| \mathrm{SRM}_{\pi(n)}|\cdots| \#}_{m \text{ slots}}
\end{equation}
Here, $\pi$ represents the permutation of identities, which remains unknown to the SP and other participants. 

\subsubsection{Collision Resolution}
In this protocol, a \textit{collision} is the event where multiple participants select the same transmission slot. Let's denote the value observed in slot $j$ as $S_j$, and the set of SRMs sent in that slot as $\{SRM_1, SRM_2, \dots, SRM_k\}$. The resulting value is:
\[
S_j = SRM_1 \oplus SRM_2 \oplus \dots \oplus SRM_k
\]
We can analyze three possible outcomes for any given slot:

\begin{itemize}
    \item Empty Slot: If no participant chooses slot $j$, the XOR sum is the identity value, zero, i,e., $S_j = 0$.
    \item Successful Reservation: If only participant $P_i$ chooses slot $j$, the resulting value is simply their own SRM, i.e., $S_j = SRM_i$.
    \item Collision (Two or More Participants): If participants $P_i$ and $P_k$ both choose slot $j$, the result is the XOR sum of their SRMs, i.e., $S_j = SRM_i \oplus SRM_k$.
\end{itemize}

These SRMs are generated to be unique, high-entropy bitstrings (akin to random numbers). When two distinct SRMs are XORed, the result is a new bitstring that is, with overwhelmingly high probability, different from both of the original SRMs. 
This makes it impossible for the colliding participants to confirm their reservation from the public data.

The collision resolving unfolds in following way:
\begin{enumerate}
    \item \textit{Publication:} SP publishes the final vector $C$ to all participants. 
    \item \textit{Private Verification:} Each participant $P_i$ scan the public vector $C$ to find a value that matches their unique, privately held $SRM_i$. If their unique $SRM_i$ is present in a slot $j$, they internally flag themselves as "successful" and record their winning slot $j$. If their $SRM_i$  is not present, they internally flag themselves as "colliding.".
    \item \textit{Automated Rerun:}  All $n$ participants establish fresh keys $S_{ij}$ and generate new SRMs. Successful Participants resubmit their new SRM to the same slot they previously won. Colliding Participants choose a new random slot from the set of slots that contained `0` in the previous public vector, $C$.
    \item \textit{Termination Condition:} Publication, Private Verification and Automated Rerun step continues until round $r$ and the protocol terminates when the server observes a public vector $C_r$ where the number of occupied slots is exactly equal to $n$.
\end{enumerate}
This procedure ensures that an external adversary sees $n$ participants submitting messages in every round, making it impossible to distinguish successful users from colliding users based on traffic patterns. The security of this stage is analyzed in detail in Section \ref{sec:slot_reservation_analysis}.

Once the iterative resolution process concludes, a final vector $C$ with $m$ exists where $n$ slots are filled with unique SRMs and $m-n$ slots are empty. To finalize the assignments, the SP performs a consolidation step, where the SP scans the vector $C$ and filters out all empty slots.  It then creates a new, compact vector, $C_{\text{final}}$, of $n$ slots consisting only of the $n$ successful SRMs, ordered by their original slot position in $C$. The SP publishes this definitive vector $C_{\text{final}}$. Each participant $P_i$ finds their most recently used SRM in this vector. Their index in $C_{\text{final}}$, now becomes their official slot position, $\mathit{pos}_i$, for the subsequent data transmission stage.

\subsection{Data Submission Stage}
Using a QKD protocol, every pair of participants $(P_i, P_j)$ establishes a shared secret seed $S_{ij}$. These seeds will be used to generate cryptographic one-time pads. Each participant $P_i$ individually prepares their data for submission. This involves several steps:
\begin{enumerate}
    \item  $P_i$ constructs a bitstream $C_{i}^{\text{msg}}$ of a total length $L_v$ (where $L_v$ is large enough to hold all $n$ messages).
    \begin{equation}
        C_{i}^{\mathrm{msg}} = \underbrace{\overbrace{00 \cdots 0}^{L_{b}}|\mathrm{msg}_{i}|00 \cdots 0}_{L_{v}}
        \label{Ci}
    \end{equation}
    This vector contains the participant's message, $\mathrm{msg}_i$, in its designated slot $\mathit{pos}_i$, with all other bits set to zero. 
    \item  Using its $n-1$ pairwise shared QKD seeds, $P_i$ generates $n-1$ pseudo-random bitstreams (pads) $C_{ij}$, each of length $L_v$. These are generated using a Quantum-Secure Pseudo-Random Function (QS-PRF).
    \begin{equation}
        C_{ij} = \mathrm{QS\text{-}PRF}\{L_{v}, S_{ij}\}, \quad \text{for } j \in \{1,\cdots, n \}, j \neq i
    \end{equation}
    \item  $P_i$ computes its final vector $C_i$, as
    \begin{equation}
        C_{i} = C_{i}^{\text{msg}} \oplus \bigoplus_{j \neq i} C_{ij}
        \label{eq:vectorCi}
    \end{equation}
    by XORing its message vector with all the generated pads. 
\end{enumerate}
Each participant $P_i$ then transmits only their final, masked vector $C_i$ to the Service Provider. The SP receives all $n$ vectors and performs a single operation: it XORs all of them together. 
\begin{equation}
    C = \bigoplus_{i=1}^{n} C_{i} =\bigoplus_{i=1}^{n} C_{i}^{\text{msg}} = \mathrm{msg}_{\pi(1)} | \mathrm{msg}_{\pi(2)} | \cdots | \mathrm{msg}_{\pi(n)}
\end{equation}
The final vector $C$ is a perfect concatenation of all participant messages, but the SP cannot determine the origin of any individual message.

To ensure data integrity, the SP broadcasts the complete, concatenated vector $C$ back to all $n$ participants. This step creates a public record and allows for community-wide verification. Each participant $P_i$ is responsible for locating their own message, $\mathit{msg}_i$, within the global vector $C$, and verifying that their message is present and has not been altered or corrupted during transmission or assembly. 

Following verification, each participant signals the outcome to the SP using a simple validation flag. This flag is a single binary bit, $f_i$, defined as:
\[
f_i =
\begin{cases}
    1 & \text{if } \mathit{msg}_i \text{ is correct and complete within } C \\
    0 & \text{if } \mathit{msg}_i \text{ is incorrect/corrupted/missing from } C
\end{cases}
\]
The participants transmit their respective flags back to the SP. The SP assembles these flags into a final validation vector, $F = [f_1, f_2, \dots, f_n]$.

The SP makes a final decision based on the contents of the validation vector $F$. If all flags are 1 (i.e., $\forall i \in  \{1,\cdots, n \}, f_i = 1$), the data is deemed correct and the protocol has succeeded. The SP forwards the final data vector $C$ to the Task Requester (TR). If one or more flags are 0 (i.e., $\exists i \in  \{1,\cdots, n \} \text{ such that } f_i = 0$), an error has been detected. The SP withholds the data from the TR and initiates an error recovery protocol. This recovery step would typically involve requesting a re-transmission only from the specific participant(s) who signaled an error, making the process efficient.

\section{Security Analysis}
\label{Security Analysis}

We analyze QADR's security with respect to the goals defined in Section \ref{sec:system_model}.

\subsection{Sender Anonymity and Unlinkability}
The anonymity of the protocol rests on the properties of the BTP mechanism fortified by quantum-secure primitives. An adversary, including a malicious SP or a coalition of up to $n-2$ participants, sees only the final XOR sums ($C$ in the reservation stage and $C$ in the submission stage).

Consider an honest participant $P_i$. Their contribution $C_i$ is masked by pads $C_{ij}$ shared with every other participant $P_j$. If participant $P_j$ is also honest, the pad $C_{ij}$ is generated from a secret key $S_{ij}$ known only to them. This key is information-theoretically secure due to QKD. The pad itself is computationally indistinguishable from random for a quantum adversary due to the QS-PRF. As long as at least one other honest participant exists, the contribution of $P_i$ is perfectly masked. An adversary cannot break the XOR sum to isolate $P_i$'s original vector $C_{i}^{\text{msg}}$ and thus cannot determine which slot $P_i$ chose or which message $P_i$ sent. This provides both sender anonymity and unlinkability.

Furthermore, the protocol's design reinforces these properties. The use of fresh QKD keys for each stage and each round of reservation prevents the adversary from linking participant activities across time. However, the interactive nature of the slot reservation stage introduces specific nuances to the anonymity guarantees, which we analyze in detail in the following section.

\subsection{Security Analysis of the Slot Reservation Stage}
\label{sec:slot_reservation_analysis}
Contribution of our work is the formal security analysis of the dynamic, collision-based slot reservation architecture. While highly scalable, this approach has inherent properties that impact its anonymity guarantees, which have been previously unexamined \cite{li2021}. Our analysis establishes the precise security bounds of this mechanism.

\subsubsection{Anonymity Set Partitioning Attack:}
The protocol's vulnerability lies in the information revealed by the public vector $C^{(r)}$ at the conclusion of each round $r$. From this vector, the adversary can construct two disjoint sets of slot indices: the set of occupied slots - $\mathcal{M}_{\text{occ}}^{(r)} = \{j \mid S_j^{(r)} \neq 0\}$; and the set of empty slots - $\mathcal{M}_{\text{emp}}^{(r)} = \{j \mid S_j^{(r)} = 0\}$

The protocol rules for the \textit{Automated Rerun} step require participants to behave differently based on the outcome of the previous round. This allows the adversary to partition the initial anonymity set, $\mathcal{P} = \{P_1, \dots, P_n\}$, into two distinct subsets after the first round:

\begin{enumerate}
    \item The Set of Successful Participants, $\mathcal{P}_{\text{succ}}^{(1)}$: Participants who successfully reserved a slot. The adversary knows that in round 2, every member of this set is constrained to resubmit their new SRM to a slot $j \in \mathcal{M}_{\text{occ}}^{(1)}$.
    \item The Set of Colliding Participants, $\mathcal{P}_{\text{coll}}^{(1)}$: Participants whose initial submissions resulted in a collision. The adversary knows that in round 2, every member of this set is constrained to choose a new random slot $j \in \mathcal{M}_{\text{emp}}^{(1)}$.
\end{enumerate}

Mathematically, these two sets form a partition of the total participant set:
\[
\mathcal{P} = \mathcal{P}_{\text{succ}}^{(1)} \cup \mathcal{P}_{\text{coll}}^{(1)} \quad \text{and} \quad \mathcal{P}_{\text{succ}}^{(1)} \cap \mathcal{P}_{\text{coll}}^{(1)} = \emptyset
\]

The sizes of these sets are directly determined by the collision structure vector 
\begin{equation}
   \vec{c} = (c_1, c_2, \dots, c_n)  
\end{equation}
from the first round, where, $c_k$ denotes the number of slots containing exactly $k$ participants. That is $c_1$: Counts the number of slots with just 1 participant (these are the successful, non-colliding participants), $c_2$: Counts the number of slots with exactly 2 participants (these are the pairwise collisions), $c_3$: Counts the number of slots with exactly 3 participants (triple collisions), and so on.  This vector must satisfy the constraint that all participants are accounted for, $\sum_{k=1}^{n} k \cdot c_k = n$. The total number of occupied slots for a given structure is $j = \sum_{k=1}^{n} c_k$. The number of successful participants is $|\mathcal{P}_{\text{succ}}^{(1)}| = c_1$, and the number of colliding participants is $|\mathcal{P}_{\text{coll}}^{(1)}| = \sum_{k=2}^{n} k \cdot c_k = n - c_1$.

\subsubsection{Quantifying Anonymity and Proving Unlinkability}
Our analysis quantifies the precise impact of this partitioning. Initially, the probability of correctly guessing the author of an SRM is $1/n$. After the first round, if an adversary observes an action associated with the colliding group, the probability of identifying a specific actor within that group increases to $1/|\mathcal{P}_{\text{coll}}^{(1)}| = 1/(n-c_1)$.

Crucially, our work proves that this property, while reducing the anonymity set size, does not break the protocol's core security guarantee: unlinkability is rigorously maintained within each partition. For a $k$-way collision, an adversary observes only the garbled sum $S_j = \bigoplus_{i=1}^{k} \text{SRM}_i$. Due to the properties of XOR and the high entropy of the SRMs, the adversary cannot recover the original SRMs, preserving $k$-anonymity for the participants in that collision.

To provide a more formal measure of the anonymity reduction, we can extend our analysis using Shannon entropy, which quantifies an adversary's uncertainty about the sender's identity.

Initially, before any information is revealed, the adversary must consider all $n$ participants as equally likely senders. The initial entropy, representing maximum uncertainty and thus the strongest anonymity, is given by:
\begin{equation}
    H_{\text{initial}} = -\sum_{i=1}^{n} \frac{1}{n} \log_2\left(\frac{1}{n}\right) = \log_2(n) \text{ bits}
\end{equation}
After the first round, if an adversary observes an action originating from the colliding group $\mathcal{P}_{\text{coll}}^{(1)}$ of size $n - c_1$, their uncertainty is reduced. The entropy of their knowledge, now confined to this smaller set, becomes:
\begin{equation}
    H_{\text{final}} = \log_2(|\mathcal{P}_{\text{coll}}^{(1)}|) = \log_2(n - c_1) \text{ bits}
\end{equation}
The information gain for the adversary, which corresponds directly to the loss of anonymity, is the reduction in entropy. This measures precisely how many bits of information the leak provides:
\begin{align}
    I_{\text{gain}} = & H_{\text{initial}} - H_{\text{final}}  = \log_2(n) - \log_2(n - c_1) \\ \nonumber = & \log_2\left(\frac{n}{n - c_1}\right) \text{ bits}
\end{align}
This information-theoretic analysis formalizes our earlier quantification. It confirms that while unlinkability is preserved, the partitioning provides a measurable amount of information to the adversary, establishing the precise bounds of the anonymity provided by the reservation stage.

It is crucial to acknowledge the practical security implications of this leak. While the protocol is secure against a passive adversary, a sophisticated active adversary controlling a large coalition of $n-k$ participants could potentially exploit this partitioning. For instance, a malicious coalition could deliberately engineer collisions to force the $k$ honest participants into the $\mathcal{P}_{\text{coll}}^{(1)}$ set. If collisions persist over subsequent rounds, the size of this anonymity set of colliding users continues to shrink, further increasing the adversary's probability of identifying a specific target within that diminishing group. This highlights that the protocol's high scalability comes at the cost of a measurable reduction in anonymity against strong, coordinated adversaries.

\subsection{Mitigating Anonymity Set Partitioning with a Verifiable Oblivious Shuffle}
\label{Verifiable OS}
The information leak described is a direct consequence of a deliberate design trade-off to maximize scalability and performance. For applications demanding stronger protection against the active adversaries detailed above, the iterative, collision-based reservation mechanism can be replaced with a single-round cryptographic protocol based on a quantum resistant verifiable oblivious shuffle \cite{Benhamouda21}. This creates a variant of the protocol with enhanced security guarantees. The modified reservation stage would proceed as follows:
\begin{enumerate}
    \item  Each participant $P_i$ encrypts a unique, random pseudonym $PN_i$ using the Service Provider's (SP) public key, which is based on a quantum-resistant asymmetric scheme (e.g., CRYSTALS-Kyber).
    \item  The SP collects all $n$ encrypted pseudonyms. It then performs a verifiable oblivious shuffle, which permutes the ciphertexts and generates a zero-knowledge proof $\pi$ that the shuffle was performed correctly without altering, adding, or removing any submissions.
    \item The SP decrypts the shuffled list of ciphertexts and publishes the resulting permuted list of pseudonyms, $L$, along with the proof $\pi$. Each participant $P_i$ privately finds their pseudonym in $L$ to determine their assigned slot.
\end{enumerate}
This framework is specifically designed to mitigate the  threat model \ref{ThreatModel}. The core defense lies in the protocol's use of a quantum-resistant cryptosystem for the initial submissions. An adversary who harvests these encrypted pseudonyms will find them resistant to decryption even by a future quantum computer. This ensures the long-term confidentiality of the submissions and preserves the unlinkability between a participant and their assigned slot, directly thwarting the adversary's goal of retroactive deanonymization.

This approach robustly mitigates the leak. Since all participants follow an identical, non-iterative procedure, there are no distinct behavioral groups for an adversary to observe, thus preventing the partitioning of the anonymity set. The verifiable proof $\pi$ ensures the integrity of the process against a malicious SP. The primary trade-off is a shift in complexity: this method introduces a higher computational burden on the SP and requires a PQC public-key infrastructure, but in exchange, it provides stronger, provable anonymity for the reservation stage.

\subsection{Adversary Model Considerations}
The choice of a centralized, untrusted Service Provider (SP) in the QADR architecture is a strategic design decision that enhances scalability and simplifies the security model compared to fully-connected peer-to-peer (P2P) systems like APMT. In P2P DC-nets, communication complexity is high ($O(n^2)$ connections), and collusion analysis is complicated by the network topology; colluding neighbors can potentially isolate and deanonymize a user~\cite{wacksman2004}. In contrast, QADR's star topology simplifies communication to $O(n)$ connections between participants and the SP. The SP acts as a simple, stateless aggregator, performing only XOR operations. The security analysis is thus simplified: as long as at least two honest participants exist, the XOR sum remains secure against the SP and upto $n-2$ colluding participants. The SP is architecturally a component for efficiency, not a trusted third party for security. This security model, which guarantees privacy as long as at least two participants remain honest, is inherited from the foundational principles of DC-Nets. It ensures the protocol is resilient even against a powerful active adversary that has compromised all but two participants, highlighting the significant robustness of the solution.

\section{Performance Evaluation}
\label{sec:performance}

\subsection{Collision Resolution Analysis}

\subsubsection{Analytical Model for Multi-Round Collision Probability}
\label{Collison_Probablility}
The slot reservation process, where $n$ participants select from $m$ available slots, is analogous to the classic "balls and bins" problem ~\cite{mitzenmacher2005}. The probability of at least one collision, $P_{m,n}$ is ~\cite{feller1968}:
\begin{equation}
P_{m,n} = 1 - P(\text{no collision}) = 1 - \prod_{k=1}^{n-1} \left(1 - \frac{k}{m}\right)
\end{equation}
While the probability of at least one collision is well-understood, a more granular analysis is required to evaluate multi-round scenarios where participants involved in collisions must try again. To this end, we derive a comprehensive analytical formula to compute the probability of any specific collision structure. 

The probability of observing a specific collision structure $\vec{c}$ in a single round, with a total sample space of $m^n$ possible outcomes, is given by the general formula (see appendix \ref{Analytical_Derivation}):

\begin{equation}
P(\vec{c}) = \frac{1}{m^n} \cdot \frac{n! \cdot m!}{(m-j)! \cdot \prod_{k=1}^{n} (k!)^{c_k} c_k!}
\end{equation}

This formula is derived by first counting the number of ways to partition $n$ participants according to the structure $\vec{c}$ and then counting the number of ways to assign these distinct participant groups to $j$ distinct slots.

This framework extends to subsequent rounds by conditioning on the outcome of the preceding round. According to the collision resolution protocol, the $c_1$ participants who successfully secured a slot alone do not participate further. Therefore, the parameters for the subsequent round are updated to $n' = n - c_1$ participants and $m' = m - j$ available slots. The probability of a new structure $\vec{c'}$ in the second round is the conditional probability $P(\vec{c'}|\vec{c})$, which is calculated using the same general formula with the new parameters $n'$ and $m'$ (see appendix \ref{Analytical_Derivation}).

The joint probability of a specific sequence of outcomes across multiple rounds is the product of the probabilities of each round. For instance, with $n=5$ participants and $m=10$ slots, the probability of observing exactly one pair and three successful singles in the first round is 50.4\%. Subsequently, the conditional probability that the two colliding participants secure unique slots in the second round (with parameters $n'=2, m'=7$) is $\approx 85.7\%$. The total joint probability of this two-round sequence is therefore $0.504 \times 0.857 \approx 0.432$.

This analytical model provides a precise and efficient method for calculating the probability of any sequence of collision outcomes.

\subsubsection{Simulation for Collision Resolution Analysis}
\begin{figure}[!t]
    \centering
    \includegraphics[width=\columnwidth]{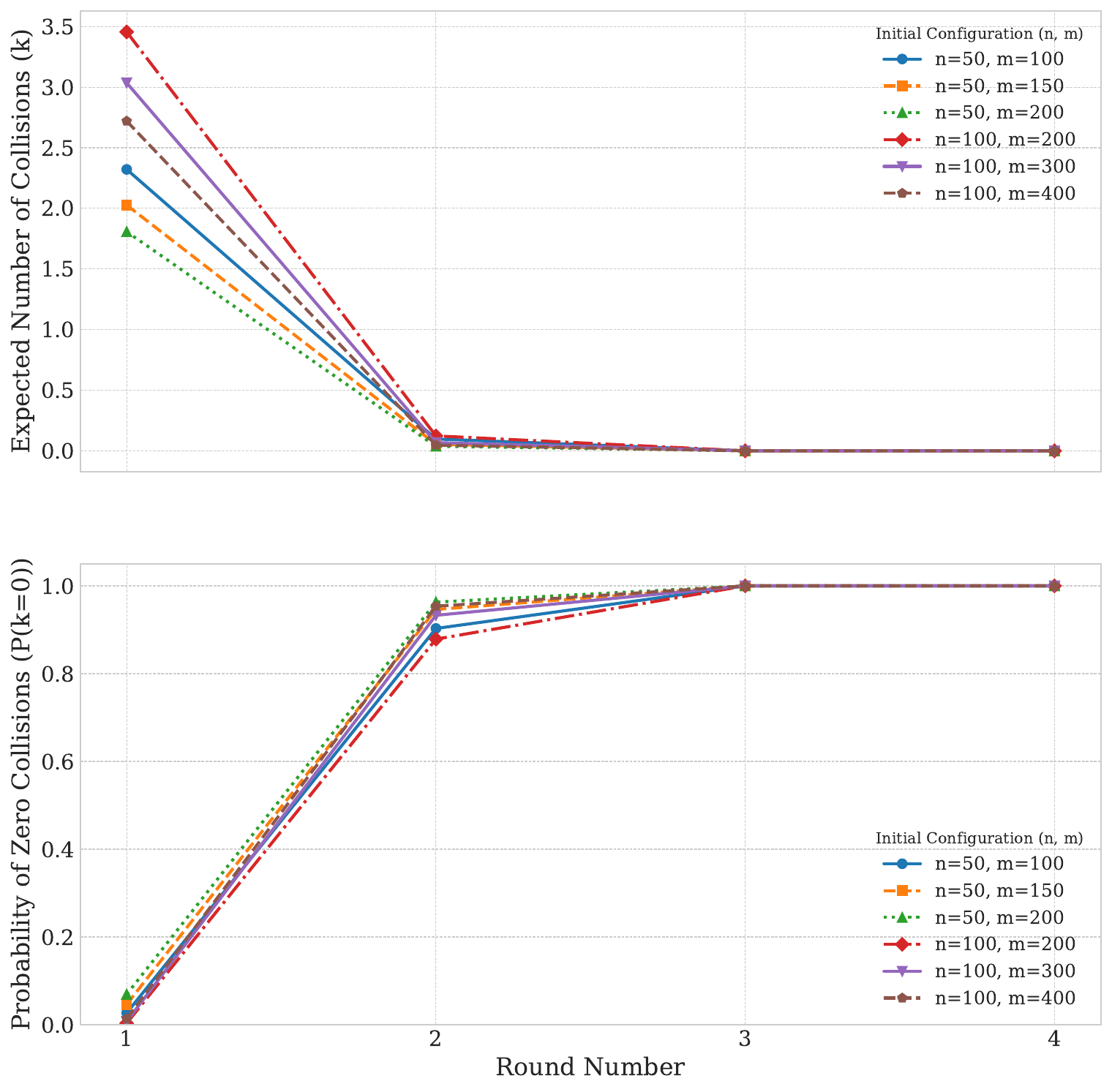}
    \caption{Simulation results showing the average number of collisions per round (top) and the probability of full resolution (bottom). A higher $\gamma$ leads to faster resolution.}
    \label{fig:collision_analysis}
\end{figure}

\begin{figure}[h]
    \includegraphics[width=\columnwidth]{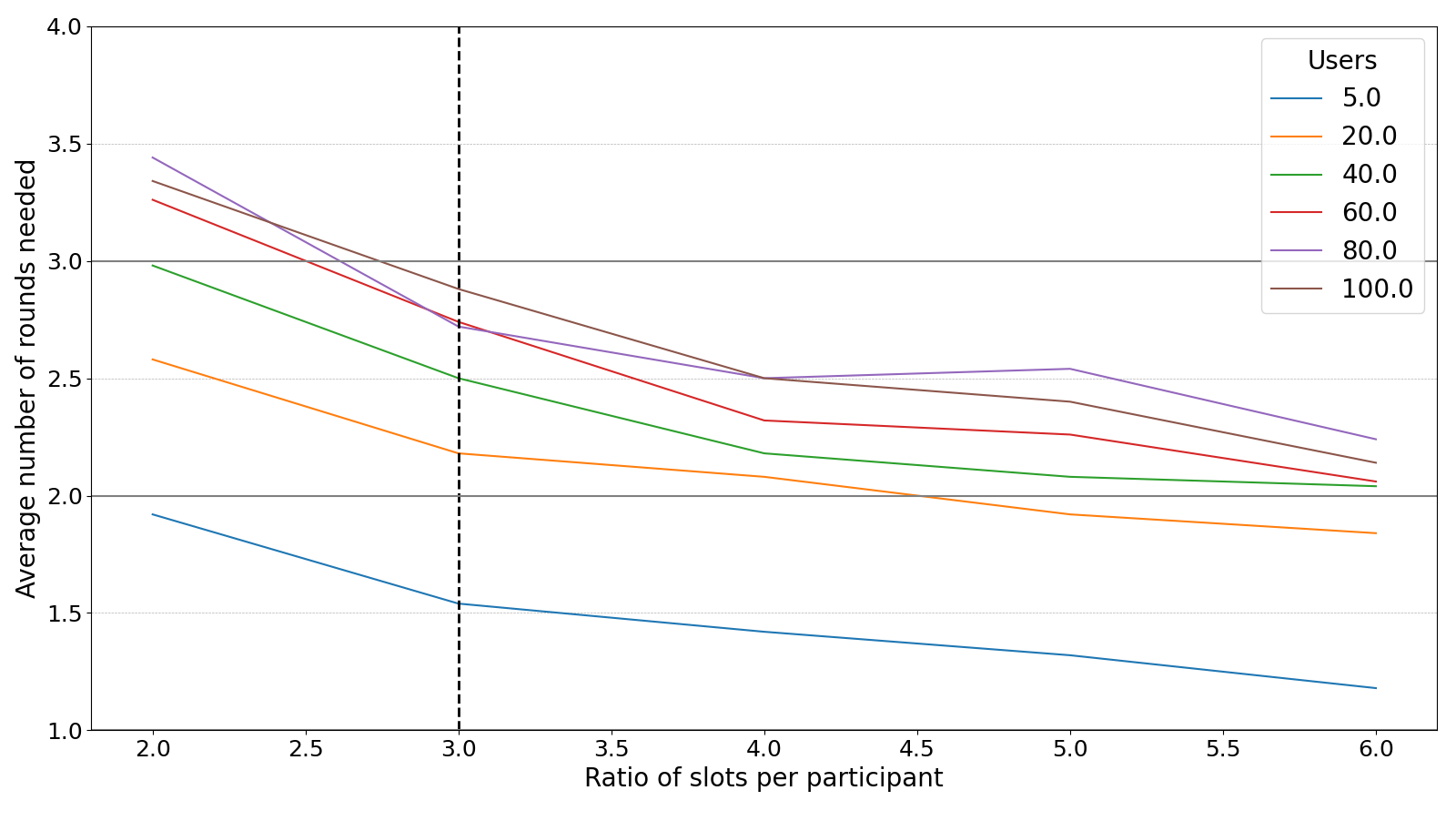}
    \caption{To determine optimal parameters, the slot reservation phase was simulated 1000 times. Our analysis identifies a slot-to-participant ratio of $\gamma=3$ as the most efficient configuration, requiring an average of $r=3$ rounds for successful resolution. These values are therefore adopted for all subsequent cost calculations.}
    \label{fig:optimal_parameter}
\end{figure}

To empirically ground our analysis, we modeled the iterative slot reservation phase to determine the relationship between the slot-to-participant ratio ($\gamma = m/n$) and the number of rounds ($r$) required for successful resolution. We implemented the analytical model  in Python, to calculate the expected number of collisions and the probability of achieving a collision-free state ($P(k=0)$) in each successive round for various configurations of $n$ and $m$. The results, visualized in Figure~\ref{fig:collision_analysis} and \ref{fig:optimal_parameter}, illustrate a clear trade-off: higher values of $\gamma$  reduce the initial number of collisions and lead to faster resolution, but at the cost of a larger, more expensive slot vector. Conversely, a lower ratio (e.g., $\gamma=2$) is less costly per round but requires more rounds to resolve all collisions. Our analysis identifies a ratio of $\gamma=3$ as the most efficient configuration, striking an optimal balance by reliably resolving all collisions within an average of three rounds, see Figure~\ref{fig:optimal_parameter}. Therefore, we adopt $\gamma=3$ and $r=3$ as the standard parameters for our subsequent cost calculations, representing a practical and resource-efficient choice.

\begin{figure}[h]
    \includegraphics[width=\columnwidth]{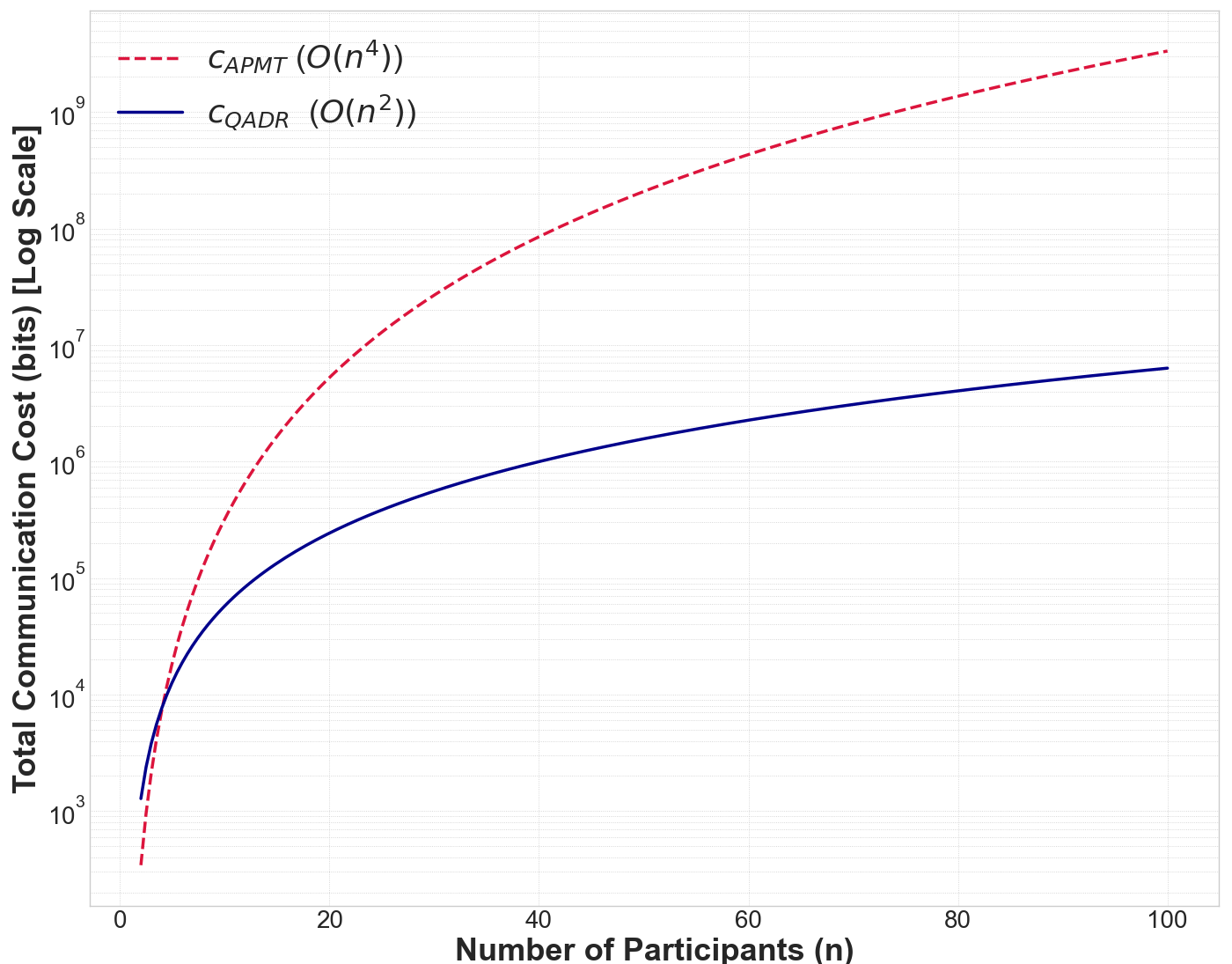}
    \caption{Plot comparing the resource cost scaling of APMT ($O(n^4)$) and QADR ($O(n^2)$) for N messages. Parameters: $\beta=16$, $\lambda = 256$ bits, $l_{\text{mes}} = 1024$ bits, $\gamma=3$, and an average of $r=3$ rounds}
    \label{fig:cost_comparison}
\end{figure}

\subsection{Cost Analysis}
We analyze the communication cost of each protocol in terms of the total number of secret bits consumed. The fundamental difference lies in their design: APMT is a serial protocol designed for a single sender \cite{broadbent2007}, whereas QADR is a parallel protocol allowing all $n$ participants to transmit simultaneously.

In APMT protocol, to send a single message of length $m$, the sender must perform collision detection, notification, message transmission, and a final veto round for acknowledgment. The cost for a single message transmission using the APMT protocol, $c_{\text{APMT}}$, for  a security parameter $\beta$  (e.g., $\beta=16$), is given by \cite{Huang2022}:
\begin{equation}
c_{\text{APMT}} = 2\beta n^2(n-1) + \left(m+2(\log_2[m+1]+\beta)\right)\frac{n(n-1)}{2}
\label{eq:APMT_cost}
\end{equation}
To compare this with QADR, we must consider the effective cost for transmitting $n$ messages (one from each participant), which requires $n$ sequential runs of the protocol, yielding a total cost of $n \times c_{\text{APMT}}$. This serial dependency, dominated by the collision detection and veto sub-protocols, results in a comparative cost complexity of $O(n^4)$.

In contrast, QADR is a parallel protocol. Its total cost all $n$ participants to send their messages in parallel is the sum of costs for the reservation and submission stages. Since each of the $n$ participants sends their vector to the central server in our star topology, the total number of vectors transmitted per round is $n$. The total communication bandwidth is:
\begin{align}
\text{BW}_{\text{QADR}} = \underbrace{r \cdot n \cdot (\gamma n \cdot l_{\text{srm}})}_{\text{Slot Reservation}} + \underbrace{n \cdot (n \cdot l_{\text{mes}})}_{\text{Data Submission}}
\label{eq:qadr_bw_cost}
\end{align}
The bandwidth cost is dominated by terms proportional to $n^2$, resulting in a complexity of $O(n^2)$. The QKD key consumption uses a short seed of length $\lambda$ (e.g., 256 bits) for the QS-PRF. Fresh keys are required for the setup stage, each of the $r$ reservation rounds, and the final data submission stage, for a total of $r+2$ key exchanges.
\begin{align}
\text{c}_{\text{QADR}} = (r+2) \cdot \lambda \cdot \frac{n(n-1)}{2}
\label{eq:qadr_key_cost}
\end{align}
This cost also scales as $O(n^2)$, but with a much smaller constant factor ($r, \lambda$) compared to the bandwidth. This highlights the efficiency of the hybrid approach in conserving the most valuable resource.


To illustrate the practical implications of this complexity difference, Figure~\ref{fig:cost_comparison} plots the costs using typical parameters: $\beta=16$, $l_{\text{srm}} = 256$ bits, $l_{\text{mes}} = 1024$ bits, $\gamma=3$, and an average of $r=3$ rounds. The plot demonstrates that while the costs may be comparable for very small networks, the $O(n^4)$ scaling of APMT's effective cost makes it prohibitively expensive for any realistic participatory sensing scenario. QADR's $O(n^2)$ scaling remains practical as the number of participants grows, confirming its superior scalability for large-scale, multi-user applications.

\subsubsection{Latency Analysis}
Beyond communication bandwidth, the total time-to-completion, or latency, is a crucial metric for practical performance. While QADR's slot reservation phase is iterative, its latency characteristics are highly favorable for a scalable system. The total reservation latency is approximately $r \times \tau$, where $r$ is the number of resolution rounds and $\tau$ is the network round-trip time. Our analysis in Figure \ref{fig:optimal_parameter}, empirically demonstrates that for an optimal slot ratio ($\gamma=3$), $r$ converges to a small, constant average value (e.g., $r \approx 3$) regardless of the number of participants, $n$. Therefore, QADR's setup latency is effectively constant, or $O(1)$, with respect to network size.

This stands in stark contrast to the serial nature of APMT. To achieve the same bulk reporting task of $n$ messages, APMT must be executed sequentially $n$ times. Since each execution involves multiple interactive rounds, the total latency for the equivalent task in APMT necessarily scales linearly with the number of participants, i.e., $O(n)$. Consequently, QADR's parallel architecture is fundamentally superior for large-scale scenarios not only in its communication bit complexity but also in its total time-to-completion.

While this analysis establishes the theoretical scalability of QADR's latency, we aim to quantify these practical overheads in future work. An experimental demonstration on a multi-node QKD testbed would allow for a precise measurement of the round-trip times and total time-to-completion, providing empirical validation of the protocol's real-world performance advantages.

\subsubsection{Cost Implications of the Verifiable Shuffle Mitigation}

The high-security reservation mechanism proposed in Section \ref{Verifiable OS} significantly alters the cost profile of the protocol's setup phase, presenting a clear performance-security trade-off.

The iterative rounds of the default reservation are replaced by a single round. In this round, each of the $n$ participants sends one PQC ciphertext to the SP. The SP then broadcasts a single message containing the list of $n$ pseudonyms and a verifiable proof $\pi$. The size of this proof is the dominant factor, but this single-round communication can be more efficient in terms of latency and potentially bandwidth than the multi-round collision-based approach.

The primary cost is shifted from communication to computation. The default protocol relies on computationally trivial XOR operations. The shuffle-based alternative, however, imposes a well-defined computational load. Using state-of-the-art, post-quantum shuffle arguments \cite{Benhamouda21}, the Service Provider's cost to generate the proof is quasi-linear, scaling at $O(n \log n)$. This proof can then be verified by all participants with a more efficient linear complexity of $O(n)$. The initial cost for each participant to submit their pseudonym remains a single PQC encryption.

A notable benefit of this alternative is a reduction in the demand for quantum resources. The shuffle mechanism does not use the pairwise QKD keys ($S_{ij}$). Therefore, the number of required QKD key establishment sessions is reduced from $(r+2)$ to just two (one for the initial setup and one for the final data submission stage). The total QKD key bits consumed would be revised to:

$$KeyBits_{QADR-Shuffle} = 2 \cdot \lambda \cdot \frac{n(n-1)}{2}$$

This significantly lowers the burden on the quantum network compared to the default protocol's requirements.

\section{Conclusion and Future Research}
\label{Conclusion}
This paper has addressed the critical need for private and quantum-resistant data reporting in participatory sensing. We proposed the Quantum Anonymous Data Reporting (QADR) protocol, a novel framework that integrates the information-theoretic security of Quantum Key Distribution (QKD) with the efficiency of a Bulk Transfer Protocol (BTP). Our analysis demonstrates that QADR offers robust anonymity against both classical and quantum adversaries. A comparative evaluation reveals that QADR is orders of magnitude more scalable than existing quantum protocols like APMT, establishing its theoretical viability and superior scalability, positioning it as a high-performance goal for future quantum-secured anonymity systems deployed in real-world, large-scale applications.

While the theoretical foundation of QADR is strong, practical implementation requires further research. The protocol's primary vulnerability is to jamming attacks, a known weakness of DC-net based systems. Although QADR can detect such disruptions, future iterations should integrate lightweight, quantum-secure accountability mechanisms, such as verifiable DC-nets \cite{chaum1988dc}, to identify and exclude malicious participants. The most significant practical barrier is the current state of quantum network infrastructure \cite{qkd_future}. The limitations in QKD technology---including key generation rate, distance, and cost---make a fully-connected mobile network currently infeasible. Future efforts must focus on co-designing protocols that are tolerant of realistic, imperfect quantum networks. This could involve hybrid models that use post-quantum classical cryptography for less sensitive stages, or architectures based on trusted nodes \cite{trusted_node}.

To bridge the gap between theory and deployment, the performance and security claims of QADR must be validated through large-scale simulations and experimental implementation on a multi-node QKD testbed. Successfully addressing these challenges will be essential to paving the way for a future where large-scale data collaboration can proceed securely, privately, and with long-term integrity.

\bibliographystyle{IEEEtran}
\bibliography{biblio}

@article{li2021,
  author    = {Yang Li and Hongtao Song and Yunlong Zhao and Nianmin Yao and Nianbin Wang},
  title     = {Anonymous Data Reporting Strategy with Dynamic Incentive Mechanism for Participatory Sensing},
  journal   = {Security and Communication Networks},
  volume    = {2021},
  pages     = {1-20},
  year      = {2021},
  publisher = {Hindawi}
}

@article{broadbent2007,
  author    = {Anne Broadbent and Alain Tapp},
  title     = {Information-theoretic Security without an Honest Majority},
  journal   = {arXiv preprint},
  volume    = {arXiv:0706.2010},
  year      = {2007}
}

@article{tor,
  author={Dingledine, Roger and Mathewson, Nick and Syverson, Paul},
  title={Tor: The second-generation onion router},
  journal={Proc. 13th USENIX Security Symp.},
  volume={13},
  pages={303--320},
  year={2004}
}

@article{chaum1981,
  author={Chaum, David L.},
  title={Untraceable electronic mail, return addresses, and digital pseudonyms},
  journal={Communications of the ACM},
  volume={24},
  number={2},
  pages={84--90},
  year={1981}
}

@inproceedings{chaum1988dc,
  author={Chaum, David},
  title={The dining cryptographers problem: Unconditional sender and recipient untraceability},
  booktitle={Journal of Cryptology},
  volume={1},
  pages={65--75},
  year={1988}
}

@inproceedings{wacksman2004,
  author={Wacksman, Adam},
  title={A survey of the dining cryptographers problem},
  booktitle={Yale University Computer Science Department},
  year={2004}
}

@misc{Madqci,
      title={MadQCI: a heterogeneous and scalable SDN QKD network deployed in production facilities}, 
      author={V. Martin and J. P. Brito and L. Ortiz and R. B. Mendez and J. S. Buruaga and R. J. Vicente and A. Sebastián-Lombraña and D. Rincon and F. Perez and C. Sanchez and M. Peev and H. H. Brunner and F. Fung and A. Poppe and F. Fröwis and A. J. Shields and R. I. Woodward and H. Griesser and S. Roehrich and F. De La Iglesia and C. Abellan and M. Hentschel and J. M. Rivas-Moscoso and A. Pastor and J. Folgueira and D. R. Lopez},
      year={2023},
      eprint={2311.12791},
      archivePrefix={arXiv},
      primaryClass={quant-ph},
      url={https://arxiv.org/abs/2311.12791}, 
}

@Article{Huang2022,
author={Huang, Zixin
and Joshi, Siddarth Koduru
and Aktas, Djeylan
and Lupo, Cosmo
and Quintavalle, Armanda O.
and Venkatachalam, Natarajan
and Wengerowsky, S{\"o}ren
and Lon{\v{c}}ari{\'{c}}, Martin
and Neumann, Sebastian Philipp
and Liu, Bo
and Samec, {\v{Z}}eljko
and Kling, Laurent
and Stip{\v{c}}evi{\'{c}}, Mario
and Ursin, Rupert
and Rarity, John G.},
title={Experimental implementation of secure anonymous protocols on an eight-user quantum key distribution network},
journal={npj Quantum Information},
year={2022},
month={Mar},
day={07},
volume={8},
number={1},
pages={25},
issn={2056-6387},
doi={10.1038/s41534-022-00535-1},
url={https://doi.org/10.1038/s41534-022-00535-1}
}

@inproceedings{Benhamouda21,
  author    = {Foued Benhamouda and
               Firat Kaged and
               Phong Q. Nguyen and
               Alice Nitulescu},
  title     = {Shuffle Arguments for Lattices},
  booktitle = {Advances in Cryptology -- EUROCRYPT 2021},
  year      = {2021},
  pages     = {66--96},
  publisher = {Springer, Cham}
}

@inproceedings{christandl2005,
  author    = {Christandl, Matthias and Wehner, Stephanie},
  title     = {Quantum Anonymous Transmissions},
  booktitle = {Advances in Cryptology -- ASIACRYPT 2005: 11th International Conference on the Theory and Application of Cryptology and Information Security, Chennai, India, December 4-8, 2005, Proceedings},
  series    = {Lecture Notes in Computer Science},
  volume    = {3788},
  editor    = {Roy, Bimal K.},
  publisher = {Springer},
  address   = {Berlin, Heidelberg},
  year      = {2005},
  pages     = {217--235},
  doi       = {10.1007/11593447_13}
}

@book{mitzenmacher2005,
    author={Mitzenmacher, Michael and Upfal, Eli},
    title={Probability and computing: Randomized algorithms and probabilistic analysis},
    publisher={Cambridge university press},
    year={2005}
}

@book{feller1968,
    author={Feller, William},
    title={An introduction to probability theory and its applications, Volume 1},
    publisher={John Wiley \& Sons},
    year={1968}
}

@article{qkd_future,
    author={Pirandola, Stefano and others},
    title={Advances in quantum cryptography},
    journal={Advances in Optics and Photonics},
    volume={12},
    number={4},
    pages={1012--1236},
    year={2020}
}

@article{trusted_node,
    author={Sasaki, Masahide and others},
    title={Field test of quantum key distribution in the Tokyo QKD Network},
    journal={Optics Express},
    volume={19},
    number={11},
    pages={10387--10409},
    year={2011}
}

@inproceedings{BPR12,
  author    = {Aloni Banerjee and Chris Peikert and Alon Rosen},
  title     = {Pseudorandom Functions and Lattices},
  booktitle = {Advances in Cryptology -- EUROCRYPT 2012},
  year      = {2012},
  pages     = {719--737},
  publisher = {Springer Berlin Heidelberg},
  series    = {Lecture Notes in Computer Science},
  volume    = {7237}
}

@inproceedings{Burke2006ParticipatorySensing,
  author    = {Jeff Burke and David Estrin and Mark Hansen and Andrew Parker and Nithya Ramanathan and Sasank Reddy and Mani B. Srivastava},
  title     = {Participatory Sensing},
  booktitle = {Proceedings of the 1st International Workshop on World-Sensor-Web (WSW '06)},
  year      = {2006},
  pages     = {117--134},
  publisher = {ACM},
  address   = {Boulder, Colorado, USA}
}

@article{Christin2011Privacy,
  author    = {Delphine Christin and Andreas Reinhardt and Salil S. Kanhere and Matthias Hollick},
  title     = {A Survey on Privacy in Mobile Participatory Sensing Applications},
  journal   = {The Journal of Systems and Software},
  volume    = {84},
  number    = {11},
  pages     = {1928--1946},
  year      = {2011},
  publisher = {Elsevier}
}

@inproceedings{Danezis2015Traffic,
  author    = {George Danezis and Ian Goldberg},
  title     = {Sphinx: A Compact and Provably Secure Mix Format},
  booktitle = {Proceedings of the 30th IEEE Symposium on Security and Privacy (S\&P '09)},
  year      = {2009},
  pages     = {269--283},
  publisher = {IEEE Computer Society}
}

@inproceedings{Shor1997Polynomial,
  author    = {Peter W. Shor},
  title     = {Polynomial-Time Algorithms for Prime Factorization and Discrete Logarithms on a Quantum Computer},
  booktitle = {SIAM Journal on Computing},
  volume    = {26},
  number    = {5},
  pages     = {1484--1509},
  year      = {1997},
  publisher = {SIAM}
}

@article{Bernstein2017PostQuantum,
  author    = {Daniel J. Bernstein and Tanja Lange},
  title     = {Post-quantum cryptography},
  journal   = {Nature},
  volume    = {549},
  number    = {7671},
  pages     = {188--194},
  year      = {2017},
  publisher = {Nature Publishing Group}
}

@inproceedings{Bennett2014QuantumCrypto,
  author    = {Charles H. Bennett and Gilles Brassard},
  title     = {Quantum Cryptography: Public Key Distribution and Coin Tossing},
  booktitle = {Proceedings of IEEE International Conference on Computers, Systems, and Signal Processing},
  year      = {1984},
  pages     = {175--179},
  address   = {Bangalore, India}
}

@article{AQCKA,
  title = {Anonymous Quantum Conference Key Agreement},
  author = {Hahn, Frederik and de Jong, Jarn and Pappa, Anna},
  journal = {PRX Quantum},
  volume = {1},
  issue = {2},
  pages = {020325},
  numpages = {12},
  year = {2020},
  month = {Dec},
  publisher = {American Physical Society},
  doi = {10.1103/PRXQuantum.1.020325},
  url = {https://link.aps.org/doi/10.1103/PRXQuantum.1.020325}
}

@article{grasselli2022anonymous,
  title = {Secure Anonymous Conferencing in Quantum Networks},
  author = {Grasselli, Federico and Murta, Gl\'aucia and de Jong, Jarn and Hahn, Frederik and Bru\ss{}, Dagmar and Kampermann, Hermann and Pappa, Anna},
  journal = {PRX Quantum},
  volume = {3},
  issue = {4},
  pages = {040306},
  numpages = {23},
  year = {2022},
  month = {Oct},
  publisher = {American Physical Society},
  doi = {10.1103/PRXQuantum.3.040306},
  url = {https://link.aps.org/doi/10.1103/PRXQuantum.3.040306}
}

@article{kyber,
  title={CRYSTALS-Kyber: a CCA-secure module-lattice-based KEM},
  author={Avans, Joppe and Bos, Joppe and Ducas, L{\'e}o and Kiltz, Eike and Lepoint, Tancr{\`e}de and Lyubashevsky, Vadim and Schanck, John M and Schwabe, Peter and Seiler, Gregor and Stehl{\'e}, Damien},
  journal={IACR Transactions on Cryptographic Hardware and Embedded Systems},
  pages={1--43},
  year={2021}
}

\appendices
\section{Analytical Derivation of Collision Probability}
\label{Analytical_Derivation}
This appendix provides a general analytical formula for the probability of observing a specific collision structure in a single round.

\subsection{Formal Definition}
A collision structure is represented by a vector $\vec{c} = (c_1, \dots, c_n)$, where $c_k$ is the number of slots containing exactly $k$ participants. The structure must satisfy $\sum_{k=1}^{n} k \cdot c_k = n$. The total number of occupied slots is $j = \sum_{k=1}^{n} c_k$.

\subsection{General Probability Formula}
The number of ways to achieve a structure $\vec{c}$ is found by a three-step process. First, we count the number of ways to partition the set of $n$ distinct participants into groups that match the desired collision structure (i.e., $c_1$ groups of size 1, $c_2$ groups of size 2, etc.). This is a standard combinatorial formula for partitioning a labeled set into unlabeled groups:
\begin{equation}
W_{\text{partition}}(n, \vec{c}) = \frac{n!}{\prod_{k=1}^{n} (k!)^{c_k} c_k!}
\label{eq:partition_formula}
\end{equation}
Here, the $n!$ term represents all permutations of the $n$ participants. We divide by $(k!)^{c_k}$ to remove the permutations of participants within each of the $c_k$ groups of size $k$, as the internal ordering is irrelevant. We then divide by $c_k!$ to remove the permutations among the $c_k$ groups of the same size $k$, as these groups are considered indistinguishable before being assigned to specific slots.
 
Next, we take the $j$ distinct groups of participants formed in above step and place them into the $m$ distinct slots. This involves two sub-steps:
\begin{enumerate}
    \item \textbf{Select $j$ slots} from the $m$ available slots, which can be done in $\binom{m}{j}$ ways.
    \item \textbf{Assign the $j$ distinct groups} to these $j$ chosen slots, which can be done in $j!$ ways.
\end{enumerate}
The total number of ways to select the slots and assign the groups is the product of these two numbers, which simplifies to the permutation formula $P(m,j)$:
\begin{align*}
W_{\text{placement}}(m, j) & = \binom{m}{j} \cdot j! = \frac{m!}{j!(m-j)!} \cdot j! \\
& = \frac{m!}{(m-j)!} = P(m,j)
\label{eq:placement_formula}
\end{align*}
The total number of favorable outcomes is $N_{\text{favorable}}(\vec{c}) = W_{\text{partition}} \times W_{\text{placement}}$. Dividing by the total sample space size, $m^n$, gives the general probability formula:
\begin{equation}
P(\vec{c}) = \frac{1}{m^n} \cdot \frac{n! \cdot m!}{(m-j)! \cdot \prod_{k=1}^{n} (k!)^{c_k} c_k!}
\label{eq:general_prob_formula}
\end{equation}
This formula can be applied conditionally to model multi-round scenarios, as shown in the main text's worked examples.

\subsection{Extension to Subsequent Rounds}
The probability of an outcome in a subsequent round (e.g., Round 2) is conditional on the outcome of the preceding round (Round 1). The collision resolution protocol dictates the parameters for this next stage.

Let the outcome of Round 1 be the structure $\vec{c}$. According to the protocol, successful participants (those in slots of size 1) do not participate in the next round. Their slots are now considered occupied. The parameters for Round 2 are therefore: Number of participants in Round 2 ($n' = n - c_1$): Only those who were in collisions ($k \ge 2$) participate. Number of available slots in Round 2 ($m' = m - j$): The total slots minus those occupied in first round.
        
The probability of a new structure $\vec{c}'$ in Round 2 is the conditional probability $P(\vec{c}' | \vec{c})$, which can be found by applying the general formula (\ref{eq:general_prob_formula}) to the new parameters $n'$ and $m'$.
\begin{equation}
P(\vec{c}' \text{ and } \vec{c}) = P(\vec{c}) \times P(\vec{c}' | \vec{c})
\end{equation}

\subsection{Example: A Two-Round Scenario}
Consider $n=5, m=10$.
\subsubsection{Round 1: One Pair, Three Singles}
The structure corresponds to $c_1 = 3$ (three slots with one participant) and $c_2 = 1$ (one slot with two participants), with all other $c_k=0$. Thus, the vector is $\vec{c} = (3, 1, 0, \dots)$. Constraints check: Participants: $(1 \cdot c_1) + (2 \cdot c_2) = (1 \cdot 3) + (2 \cdot 1) = 5 = n$. (Correct), Occupied Slots: $j = c_1 + c_2 = 3 + 1 = 4$. Apply the General Formula (\ref{eq:general_prob_formula}):
\begin{align*}
P(\vec{c}) &= \frac{1}{10^5} \cdot \frac{5! \cdot 10!}{(10-4)! \cdot [(1!)^3 3!] \cdot [(2!)^1 1!]} = 0.504
\end{align*}
The probability of observing exactly one pair and three singles in the first round is 50.4\%.

\subsubsection{Round 2: Two Singles (Success)}
The parameters for Round 2 are $n' = 5-3=2$ and $m' = 10-4=6$. The desired structure is $\vec{c}'=(2,0,\dots)$, with $j'=2$.
\begin{align*}
P(\vec{c}' | \vec{c}) &= \frac{1}{6^2} \cdot \frac{2! \cdot 6!}{(6-2)! \cdot [(1!)^2 2!]}  \approx 0.833
\end{align*}
The conditional probability of the two colliding participants succeeding is $\approx 83.3\%$. 

\subsubsection{Joint Probability Across Rounds}
The total probability  for our two-round scenario (one pair in Round 1, followed by success in Round 2), the joint probability is:
\begin{equation}
P(\vec{c}' \text{ and } \vec{c}) = P(\vec{c}) \times P(\vec{c}' | \vec{c}) = 0.504 \times \frac{6}{7} \approx 0.42
\end{equation}
This framework can be applied iteratively to analyze the probability of any sequence of outcomes over any number of rounds.
\end{document}